# Electrical suppression of all nonradiative recombination pathways in monolayer semiconductors

*Der-Hsien Lien[1,2,†], Shiekh Zia Uddin[1,2,†], Matthew Yeh[1,2], Matin Amani[1,2], Hyungjin Kim[1,2], Joel W. Ager III[2,3], Eli Yablonovitch[1], and Ali Javey[1,2,\*]*

[1]Electrical Engineering and Computer Sciences, University of California, Berkeley, CA 94720, United States
[2]Materials Sciences Division, Lawrence Berkeley National Laboratory, Berkeley, CA 94720, United States
[3]Materials Science and Engineering, University of California, Berkeley, CA 94720, United States

[†]These authors contributed equally
*Address correspondence to: ajavey@berkeley.edu

**Defects in conventional semiconductors substantially lower the photoluminescence (PL) quantum yield (QY), a key metric of optoelectronic performance that directly dictates the maximum device efficiency. Two-dimensional (2D) transition metal dichalcogenides (TMDCs), such as monolayer MoS$_2$, often exhibit low PL QY for as-processed samples, which has typically been attributed to a large native defect density. We show that the PL QY of as-processed MoS$_2$ and WS$_2$ monolayers reaches near-unity when they are made intrinsic by electrostatic doping, without any chemical passivation. Surprisingly, neutral exciton recombination is entirely radiative even in the presence of a high native defect density. This finding enables TMDC monolayers for optoelectronic device applications as the stringent requirement of low defect density is eased.**

Multiparticle Coulomb interactions are particularly pronounced in transition metal dichalcogenide (TMDC) monolayers, leading to a multitude of recombination pathways, each associated with the different quasiparticles produced by these interactions *(1)*. The recombination rate of excitons formed by photogenerated carriers *(2, 3)*, depends nonlinearly on the concentration. Because excitons interact with background charge to form trions *(4-8)*, the Fermi level also controls the dominant recombination pathway. Thus, both the background carrier concentration and the generation rate must be tuned to investigate the complete effect of multiparticle interactions on TMDC photoluminescence (PL) quantum yield (QY).

In this work, we simultaneously altered the photocarrier generation rate ($G$) by varying the incident pump power, and the total charge concentration (electron and hole population densities; $N$ and $P$) by varying the back-gate voltage ($V_g$) in a capacitor structure (Fig. 1A). Surprisingly, we found that all neutral excitons recombine radiatively in as-processed monolayers of MoS$_2$, resulting in near-unity QY at low generation rates. This high QY occurred despite a reported high native defect density of $10^{10}$ to $10^{13}$ cm$^{-2}$ determined by microscopy techniques *(9)*, in stark contrast to conventional inorganic semiconductors, where even a small number of defects has detrimental effects on the PL QY *(10)*. To gain additional insight into the photophysics, we also introduce a simple kinetic model involving excitons, trions, bi-excitons and free carriers for both sulfur- and selenide-based TMDCs. Notably, we previously demonstrated that bis(trifluoromethane) sulfonimide (TFSI) treatment results in near-unity PL QY in monolayer MoS$_2$ *(11)*. However, the underlying mechanism remained unknown. Here we show that the PL enhancement mechanism is electron counter-doping without chemical defect passivation; TFSI effectively acts as a Lewis acid. Beyond solving this mystery, the work here presents a practical method of brightening TMDCs for device applications, overcoming TFSI's major limitation of stability during subsequent device processing.

We first present data from a MoS$_2$ device, which contained a monolayer of MoS$_2$ encapsulated in poly(methyl methacrylate) (PMMA), along with a transferred gold electrical contact *(12)*. PMMA reduced $V_g$ hysteresis *(13, 14)*, thus enabling more stable and accurate measurements (fig. S1). [Device fabrication details are in the supplementary materials]. The $V_g$-dependence of the PL spectra of a MoS$_2$ monolayer was measured at a high generation rate of $G = 10^{18}$ cm$^{-2}$s$^{-1}$ (Fig. 1B). The peak PL intensity at $V_g = -20$ V showed a ~300-fold enhancement compared to $V_g = 0$ V (Fig. 1B). A peak energy shift of 30 meV was also observed (inset of Fig. 1B; detailed discussion of peak position is in the supplementary materials and fig. S2), which has been attributed to the trion binding energy *(5-8, 15)*. PL images of a monolayer device showed that the enhancement of PL intensity by electrostatic doping was spatially uniform (Fig. 1, C, D, and E). Calibrated PL measurements at room temperature were performed to quantitatively extract the QY as a function of $V_g$ and $G$ (Fig. 2, A and B; measurement accuracy and precision are discussed in the supplementary materials). At $V_g = 0$ V and $G = 6 \times 10^{17}$ cm$^{-2}$s$^{-1}$, the QY was 0.1% (Fig. 2B), consistent with the pristine MoS$_2$ in previous reports *(11, 16)*. The QY increased as $V_g$ decreased; at $V_g$ = -20 V and a low generation rate of $G = 6 \times 10^{15}$ cm$^{-2}$s$^{-1}$, the QY increased by 3 orders of magnitude and reached a peak value of 75% ± 10%, (expanded figure of Fig. 2B shown in fig. S3). Through application of $V_g$, the QY of a chemically-untreated MoS$_2$ monolayer approached unity (75% ± 10%), matching the response previously observed in MoS$_2$ monolayers treated with bis(trifluoromethane) sulfonimide (TFSI) *(11)*.

In the traditional ABC-model *(17)* of carrier recombination in semiconductors, defect-mediated Shockley-Read-Hall (SRH) recombination dominates at low generation rates, while Auger recombination dominates at high generation. In the SRH regime, QY increases linearly with $G$. In the Auger regime, QY decreases with a slope of -2/3 in a log-log plot *(17)*. Neither of these power laws were observed in the pump-dependent QY behavior of MoS$_2$. Instead, the observed response can be understood by considering the interaction of excitons, trions, and free carriers and their subsequent recombination channels (shown schematically in Fig. 2C, details are in the supplementary materials). Depending on the type of carriers present, the evolution of the exciton formed by photogenerated carriers varies dramatically. If the semiconductor is intrinsic, the exciton stays neutral; if the majority charge carriers are holes or electrons, the exciton may become a positive or negative trion, respectively. For the case with negative trions, the generation rate $G$ in steady state is balanced by the rates of all of the recombination channels:

$$G = \frac{n_X}{\tau_X} + \frac{n_T}{\tau_T} + C_{bx} n_X^2 \qquad (1)$$

where $n_X$ and $n_T$ ($\tau_X$ and $\tau_T$) are the neutral exciton and negative trion population densities (lifetimes), respectively, and $C_{bx}$ is the bi-exciton annihilation coefficient *(11, 18, 19)*. The exciton and trion lifetimes have radiative ($\tau_{Xr}, \tau_{Tr}$) and nonradiative ($\tau_{Xnr}, \tau_{Tnr}$) components: $\frac{1}{\tau_X} \equiv \frac{1}{\tau_{Xr}} +$ $\frac{1}{\tau_{Xnr}}, \frac{1}{\tau_T} \equiv \frac{1}{\tau_{Tr}} + \frac{1}{\tau_{Tnr}}$. The QY is then given by the ratio of the radiative recombination rate of both trions and excitons to total generation rate $G$.

$$QY = \frac{1}{G}\left(\frac{n_X}{\tau_{Xr}} + \frac{n_T}{\tau_{Tr}}\right) \tag{2}$$

The total negative charge concentration in the monolayer is given by $N = C_{ox}(V_g - V_{th})/q$, where $C_{ox}$ is the gate-oxide capacitance, $q$ is the electronic charge and $V_{th}$ is the threshold voltage. $N = n_e + n_T$, as this negative charge density arises from negative trions ($n_T$) or free electrons ($n_e$). The trion formation and dissociation rates balance in steady state and yield a law of mass-action $n_T = Tn_X n_e$, where $T$ is the trion formation coefficient (6, 20). Combining these equations, the trion concentration can be written as a function of exciton concentration $n_X$.

$$n_T = \frac{Tn_X}{1 + Tn_X}N \tag{3}$$

At low exciton density ($Tn_X \ll 1$), the trion and exciton densities are mutually proportional ($n_T \approx TNn_X$), but at high exciton density ($Tn_X \gg 1$), the trion concentration asymptotically approaches the background electron concentration ($n_T \approx N$). Numerically solving equations (1) and (3) provided exciton and trion densities for any generation rate $G$ and gate voltage $V_g$.

This model well-describes the MoS$_2$ QY data shown in Fig. 2B. By fitting the experimental data to the model, we extracted $\tau_{Tr}$ = 110 ns; $\tau_{Tnr}$ = 50 ps; $\tau_{Xr}$ = 8 ns; $C_{bx}$ = 3.5 cm$^2$s$^{-1}$. Excitons are radiative, while trions and bi-excitons are nonradiative. Specifically, the extracted trion nonradiative lifetime was nearly three orders of magnitude lower than the trion radiative lifetime, implying that the dominant trion recombination pathway is nonradiative. This difference could be the result of a defect-assisted decay, and/or a geminate Auger-like process where the electron provides the third particle required for momentum conservation. The effective trion nonradiative lifetime $\tau_{Tnr}$ reflects the combined effects of these processes. Nonradiative Auger-like recombination of trions have also been identified in doped quantum dots *(21)*. In contrast, the model does not require an exciton nonradiative lifetime to fit the data, highlighting that pure exciton recombination is entirely radiative. One possibility is that the tightly-bound, charge-neutral excitons do not interact strongly with the native charged defects. This could be an inherent trait of all excitonic systems (*e.g.*, quantum dots), as they generally exhibit high QY. At high exciton densities (*i.e.*, high generation rates), nonradiative bi-exciton annihilation dominated. At a specific generation rate $G$, $V_g$ changed the relative populations of excitons and trions, resulting in different recombination pathways dominating (Fig. 2, D and E).

As-exfoliated monolayer $MoS_2$ is electron-rich because of donor-like chalcogenide vacancies, placing the Fermi level near the conduction band *(22)*. Thus, for $V_g \geq 0$ V, the trion nonradiative recombination rate dominated for all $G$, resulting in the low observed QY (Fig. 2D). However, by applying negative $V_g$, the background electron concentration was reduced and the majority of the quasiparticles were neutral excitons (fig. S4). This change is validated by the spectral shift (Fig. 2B and S2). In this regime, the exciton radiative recombination rate dominated (Fig. 2E) and the observed QY was near-unity (75% ± 10%) at low generation rates. At high generation rates, the exciton recombination rate varied as $G^{0.5}$ because of bi-exciton annihilation (Fig. 2E). This causes a drop in the QY (Fig. 2B).

We also studied the $V_g$ and $G$ dependence of PL QY for monolayers of $WS_2$, $WSe_2$, and $MoSe_2$. The same model accurately describes these TMDCs (table S1). Similar to $MoS_2$, $WS_2$ monolayers are inherently n-doped because of sulfur vacancies and reach near-unity QY (~90%) at $V_g = -20$ V (fig. S5) *(23)*. In contrast, $WSe_2$ and $MoSe_2$ are nearly intrinsic at $V_g = 0$ V (Fig. 3, A and B, $MoSe_2$ data in fig. S6). Thus, exciton radiative recombination dominated at near-zero $V_g$ and low $G$, resulting in the highest measured QY of ~8% (Fig. 3C). Positive and negative $V_g$ moved the Fermi level closer to the conduction band and the valence band, respectively (Fig. 3B). At $V_g$

= +20 V and -20 V, the negative and positive trion nonradiative recombination rates dominated, respectively, lowering the QY. Notably, at $V_g$ = -20 V the QY increased with $G$ from $10^{18} < G < 10^{19}$ cm$^{-2}$s$^{-1}$. This unusual response is also well explained by our model. Specifically, at $G = 10^{18}$ cm$^{-2}$s$^{-1}$, the trion concentration saturated to the total charge concentration, beyond which excess excitons could not form trions (fig. S7). The neutral exciton recombination pathway then dominated, increasing the QY until bi-exciton annihilation dominated (Fig. 3D). Finally, unlike sulfur-based TMDCs, selenide-based TMDCs did not achieve near-unity QY even when their Fermi levels are near mid-gap. The reason could either be the presence of an indirect band gap near the direct band gap *(24, 25)*, or dark excitons *(26)*.

TFSI-treated monolayer MoS$_2$ was previously shown to exhibit near-unity PL QY, but the underlying enhancement mechanism was unclear *(11, 16)*. Here we show that the effect of the treatment is electron counterdoping, justified by both PL QY and time resolved PL (TRPL) measurements. TFSI-treated MoS$_2$ and an untreated sample at $V_g$ = -20 V had similar PL QY over five orders of magnitude variation in $G$ (Fig. 4A). They also had comparable TRPL decay at a low pump fluence of 1 nJ cm$^{-2}$, with respective lifetimes of 10.2 ns and 6.9 ns (Fig. 4B). Additionally, their TRPL lifetimes matched closely for all fluences (Fig. 4C, fig. S8). The Lewis acid nature of TFSI is well-established in organic chemistry *(27)*. Our results depict that TFSI acts as a Lewis acid by withdrawing electrons from (i.e., hole doping) TMDCs *via* surface charge transfer. This explanation was further validated by the observation of a threshold voltage shift in TMDC transistors after TFSI treatment, consistent with hole doping (fig. S9). Finally, the reduction of PL QY in selenide-based TMDCs after TFSI treatment *(16)* is also consistent with the gated PL measurements, and these materials being initially near intrinsic.

For the TMDCs we studied, the QYs were highest when the monolayers were intrinsic, implying that all neutral excitons radiatively recombine even in the presence of native defects.

Although near-unity QY is commonly observed in organic dye molecules *(28)*, it is uncommon in conventional inorganic semiconductors. This work establishes TMDC monolayers for optoelectronics as they can exhibit near-unity PL QY without the stringent requirement for low defect density. This electrostatic PL enhancement scheme is simple and general enough to be applied to other excitonic systems without the need for material-specific passivation techniques.

ACKNOWLEDGEMENTS

**Funding:** Material preparation, optical characterizations, and modeling were supported by the U.S. Department of Energy, Office of Science, Office of Basic Energy Sciences, Materials Sciences and Engineering Division under contract no. DE-AC02-05CH11231 within the Electronic Materials Program (KC1201). Device fabrication were supported by the Center for Energy Efficient Electronics Science (NSF Award 0939514). **Author Contributions:** D.-H.L., S.Z.U., and A.J. conceived the idea for the project and designed the experiments. D.-H.L. and S.Z.U. performed optical measurements. S.Z.U., M.Y., H.K. and M.A. fabricated devices. D.-H.L., S.Z.U., and A.J. analyzed the data. S.Z.U. and E.Y. performed analytical modeling. J.W.A. helped design optical characterization and also discussed recombination model. D.-H.L., S.Z.U., M.Y., and A.J. wrote the manuscript. All authors discussed the results and commented on the manuscript. **Competing financial interests**: The authors declare no competing financial interests. **Data and materials availability:** All data needed to evaluate the conclusions in the paper are present in the paper or the Supplementary Materials. The materials that support the findings of this study are available from the corresponding author upon reasonable request.

SUPPLEMENTARY MATERIALS
 Materials and Methods
 Supplementary Text
 Figs. S1 to S12
 Tables S1
 References (29,30)

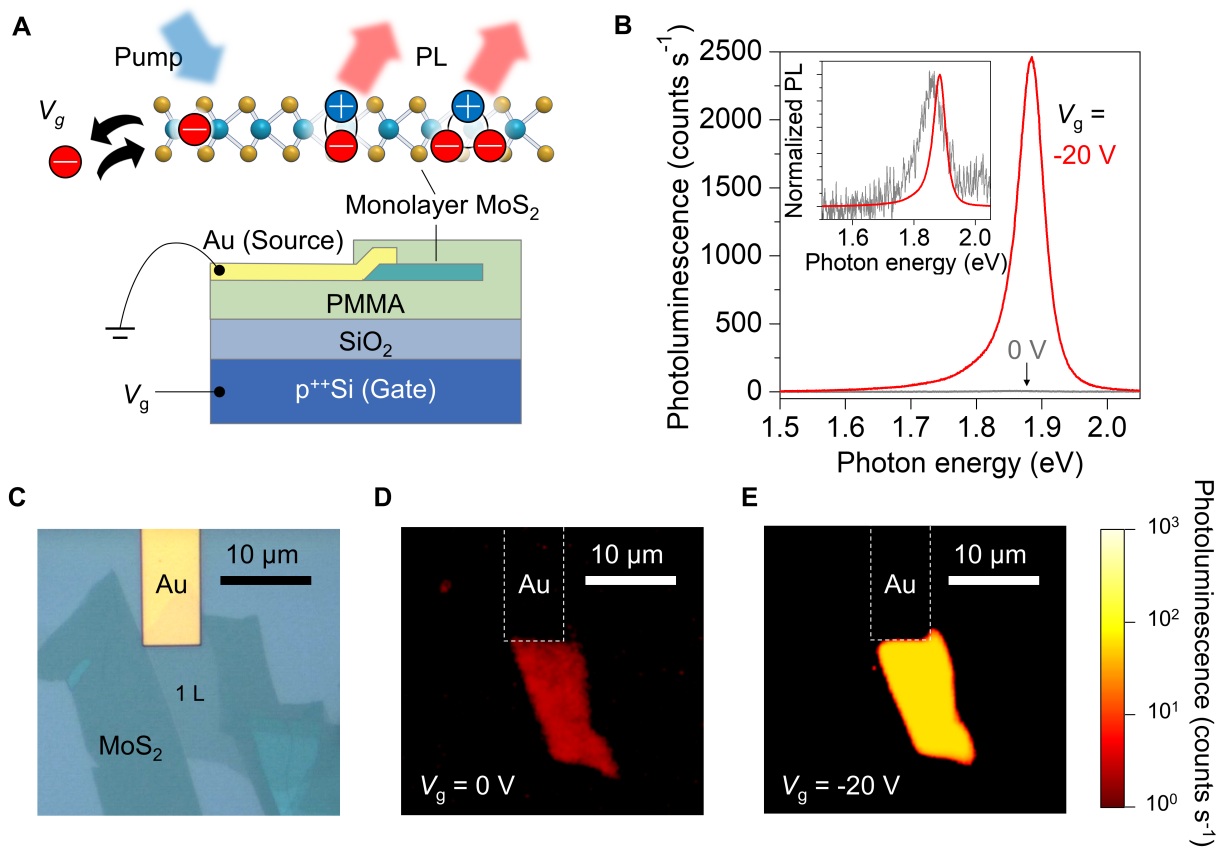

**Fig. 1. Schematics of the device and the gate dependence of photoluminescence in MoS$_2$.** (**A**) Schematic showing control of different quasiparticles by gate voltage $V_g$ and generation rate $G$. (**B**) PL spectra of the MoS$_2$ monolayer device under gate voltages $V_g$ = -20 V and 0 V at generation rate $G$ = 10$^{18}$ cm$^{-2}$s$^{-1}$. Inset is the normalized PL spectra. (**C**) Top-view optical micrograph of a MoS$_2$ device. (**D**) PL imaging of the device at $V_g$ = 0 V, and (**E**) $V_g$ = -20 V.

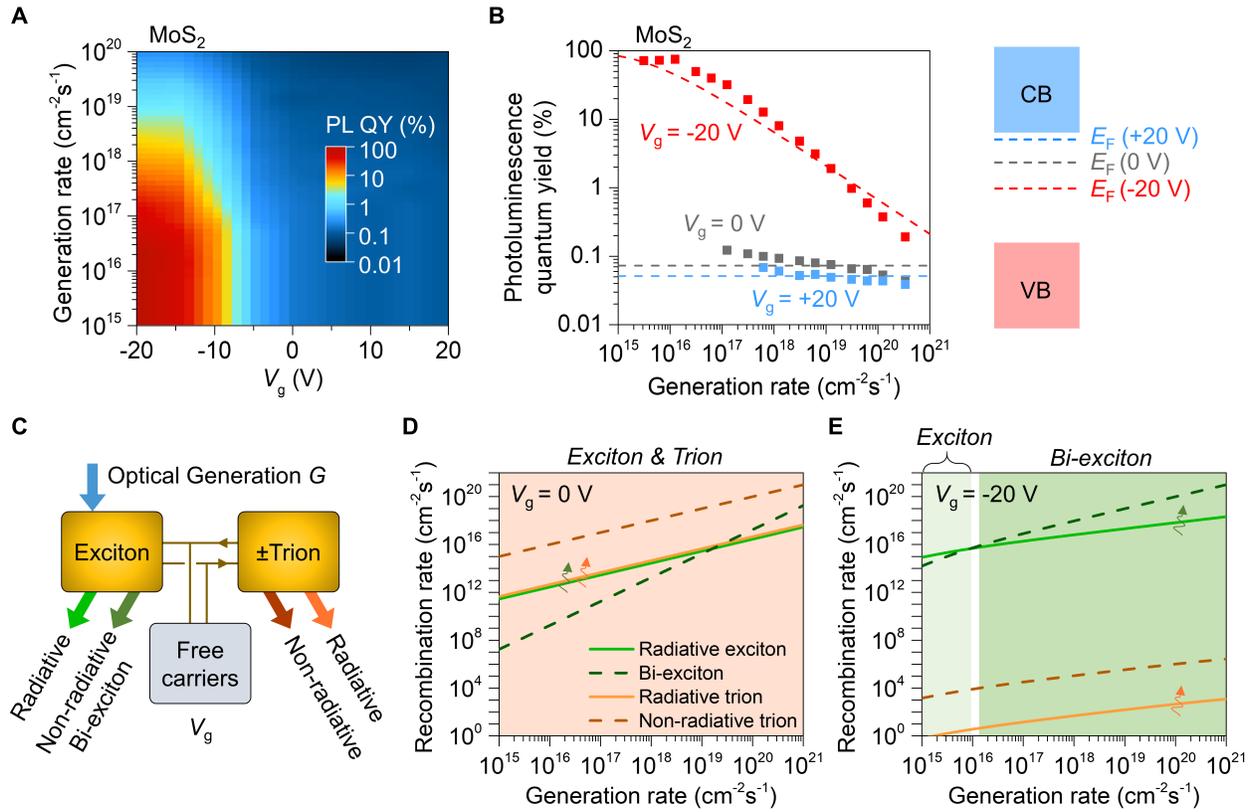

**Fig. 2. Near-unity PL QY in MoS$_2$ by electrostatic doping compensation. (A)** A color plot of MoS$_2$ PL QY vs $G$ and $V_g$. **(B)** The MoS$_2$ PL QY vs $G$ at $V_g$ = +20 V, 0 V and -20 V. Points, experimental data; dashed lines, model. Illustration of Fermi level position for different $V_g$ on the right side of the panel. **(C)** Exciton and trion recombination pathways in TMDC materials. **(D)** Calculated radiative and nonradiative recombination rates of excitons and trions in MoS$_2$ at $V_g$ = 0 V, and **(E)** -20 V.

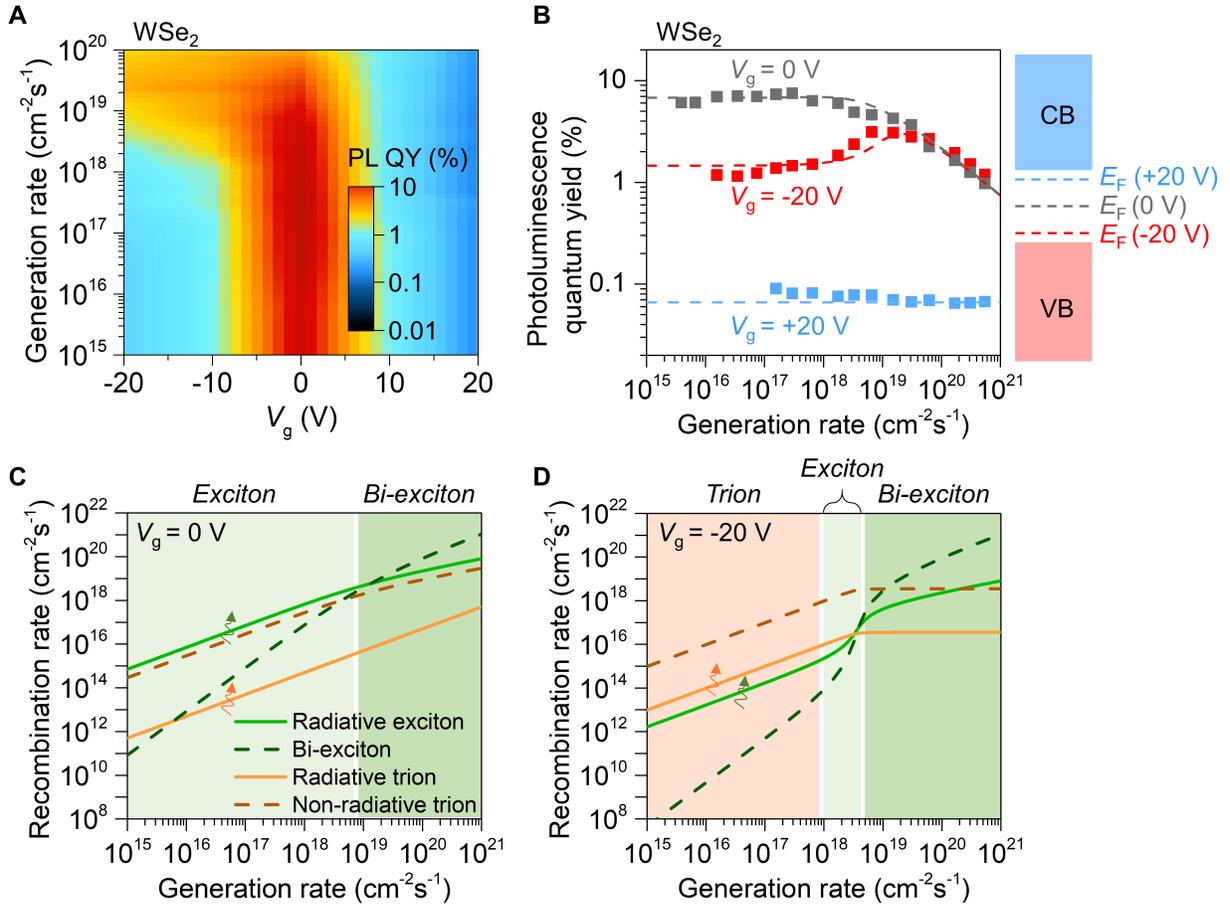

**Fig. 3. PL QY versus generation rate in WSe$_2$ monolayers.** **(A)** A color plot of PL QY vs $G$ and $V_g$ in WSe$_2$. **(B)** The PL QY vs $G$ at $V_g$ = +20 V, 0 V and -20 V. Points, experimental data; dashed lines, model. Illustration of Fermi level position for different $V_g$ on the right side of the panel. **(C)** Radiative and nonradiative recombination rates of excitons and trions in WSe$_2$ at $V_g$ = 0 V, and **(D)** -20 V. At $V_g$ = -20 V, recombination rate shows the effect of positive trion saturation leading to the observed bump in (B).

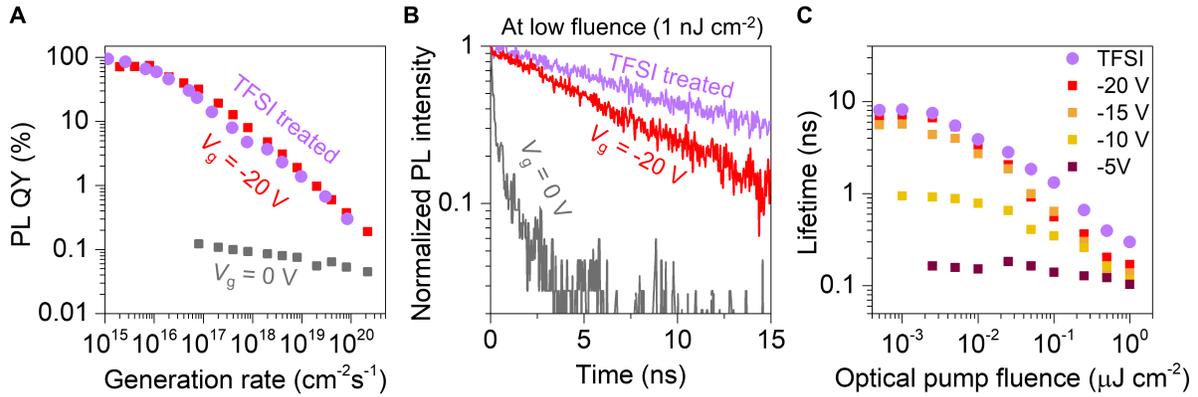

**Fig. 4. Comparison of electrostatic doping and TFSI treatment. (A)** QY vs $G$ for TFSI-treated and electrostatically-doped monolayer $MoS_2$. **(B)** TRPL of a TFSI-treated $MoS_2$ (purple), and a $MoS_2$ device under $V_g$ = 0 V (black), $V_g$ = -20 V (red) at pump fluence of 1 nJ cm$^{-2}$. **(C)** PL lifetime vs optical pump fluence for TFSI-treated $MoS_2$ and a $MoS_2$ device under various gate voltages $V_g$. Increasing $V_g$ decreases TRPL lifetime because trions have a significantly shorter lifetime than excitons. Increasing fluence decreases the lifetime through bi-exciton annihilation.